\documentclass[twocolumn,aps,prl,draft,showpacs,amssymb]{revtex4}
\usepackage{graphicx}
\usepackage{bm}
\begin{document}
\title{Energy and enstrophy fluxes in the double cascade of 2d turbulence}
\author{G.~Boffetta}
\affiliation{
Dipartimento di Fisica Generale and INFN, 
Universit\`a degli Studi di Torino, Via Pietro Giuria 1, 10125, Torino, Italy \\
and CNR-ISAC, Sezione di Torino, Corso Fiume 4, 10133 Torino, Italy}
\date{\today}
\begin{abstract}
High resolution direct numerical simulations of two-dimensional
turbulence in stationary conditions are presented. The development
of an energy-enstrophy double cascade is studied and found to
be compatible with the classical Kraichnan theory in the limit 
of extended inertial ranges.
The analysis of the joint distribution of energy and enstrophy fluxes
in physical space reveals a small value of cross correlation. 
This result supports many experimental and numerical
studies where only one cascade is generated.
\end{abstract}

\maketitle 

The existence of two quadratic inviscid invariants is the most
distinguishing feature of Navier Stokes equations in two dimensions.
On this basis, in a remarkable paper in 1967 \cite{K67},
Kraichnan predicted the double cascade scenario for two-dimensional
turbulence: an inverse cascade of kinetic energy $E=1/2 \langle v^2 \rangle$
to large scales and a direct cascade of enstrophy 
$Z=1/2 \langle \omega^2 \rangle$ to small scales 
($\omega={\bm \nabla} \times {\bm v}$ is the vorticity).
In statistically stationary conditions, when the turbulent flow
is sustained by an external forcing acting on a typical scale $\ell_{f}$
a double cascade develops.
According to the Kraichnan theory, at large scales
i.e. wavenumbers $k \ll k_{f}\sim \ell_{f}^{-1}$,
the energy spectrum has the form $E(k) \simeq \varepsilon^{2/3} k^{-5/3}$,
while at small scales, $k \gg k_{f}$, the prediction is 
$E(k) \simeq \eta^{2/3} k^{-3}$, with a possible logarithmic
correction \cite{K71}. Here $\varepsilon$ and 
$\eta \simeq k_{f}^{2} \varepsilon$ are 
respectively the energy and the enstrophy injection rate. 

Despite the importance of two-dimensional turbulence as a model
for many physical flows \cite{KM80,T02} and, more in general, 
for non-equilibrium statistical systems \cite{BBCF06}, a clear evidence 
of the coexisting two cascades on an extended range of scales
is still lacking.
The inverse energy cascade has observed in many laboratory experiments
\cite{PT97} and in numerical simulations \cite{SA81,FS84,SY93} 
with a statistical accuracy which has revealed the absence of 
intermittency corrections to the dimensional scaling \cite{BCV00}.
For what concerns the direct cascade, earlier numerical simulations and
experiments report spectra slightly steeper than $k^{-3}$ 
\cite{LSB88,KWG95}, while more recent investigations 
at high resolution are closer to Kraichnan prediction
\cite{B93,G98,LA00,PF02}.
It is important to remark that in presence of a large scale drag force 
(always present in experiments and sometimes also in numerics) one indeed
expects a correction to the classical exponent $-3$ \cite{NOAG00,BCMV02}.

Two recent experimental papers have been devoted to the study of the
double cascade \cite{R98,BK05}. Their results are substantially 
consistent with the classical scenario of Kraichnan, although 
the extension of the inertial range (in particular for the inverse
cascade) is limited.
Here we present high resolution (up to $16384^2$) direct numerical
simulations of forced 2D Navier-Stokes equations which 
reproduce with good accuracy both the cascades simultaneously.
Most of the injected energy flows to
large scales (where it is removed by friction dumping) while enstrophy
cascades to small scales (there removed by viscosity). 
We find strong numerical indications that the classical Kraichnan
scenario is recovered in the limit of two extended inertial ranges,
although we are unable to rule our the possibility of small corrections
in the direct cascade.
By looking at the two fluxes in physical space, we find a relatively
small value of the cross correlation among them. This result is
interpreted in favor of the possibility of generating a single
cascade, independently on the presence of the second inertial range.

The 2D Navier-Stokes equation for the vorticity field is
\begin{equation}
\partial_t\omega+{\bm v} \cdot {\bm \nabla} \omega =\nu\Delta\omega
-\alpha\omega + \Delta f,
\label{eq:1}
\end{equation}
where $\nu$ is the kinematic viscosity and $\alpha$ is a
linear friction coefficient (representing bottom friction or air friction)
necessary to obtain a stationary state.
The forcing term $f$ is assumed to be short correlated
in time (in order to control the injection rates) and narrow
banded in space. Specifically, we use a Gaussian forcing with correlation
function $\langle f({\bm r},t) f({\bm 0},0) \rangle = \delta(t) F
\ell_{f}^2 exp(-(r/\ell_{f})^2/2)$ in most of the simulations.
For the simulations at resolution $16384$ we use a different forcing
restricted to a narrow shell of wavenumbers.
Numerical integration of (\ref{eq:1}) is performed by pseudo-spectral,
fully dealiased, parallel code on a doubly periodic square domain 
at resolution $N^2$. Statistical quantities are computing 
by averaging over several large-scale eddy turnover times in stationary
conditions (only over a fraction of eddy turnover time for the 
$16384$ run because of limited resources). 
Table \ref{table1} reports the most important parameters of our simulations.

One of the simplest information which can be obtained from Table \ref{table1} 
is related to the energy-enstrophy balance.
At $N=2048$ only about one half of the energy injected is transfered
to large scales where it is removed by friction at a rate
$\varepsilon_{\alpha}=2 \alpha E$. 
This fraction increases with the resolution and becomes about $95~\%$ 
for the $N=16386$ run.
The remaining energy injected is dissipated by viscosity at scales 
comparable with the forcing scale and at a rate proportional to $\nu$
(which thus decreases by increasing the resolution).

On the other side, most of the enstrophy (around $90~\%$)
follows the direct cascade to small scales, where it is dissipated 
by viscosity. We observe a moderate increase of the large scale
enstrophy dissipation $\eta_{\alpha}$ increasing the resolution:
this is a finite-size effect due the increase of
$\alpha$ with $N$ (see Table~\ref{table1}) necessary to keep the friction scale 
$\ell_{\alpha}\simeq \varepsilon_{\alpha}^{1/2} \alpha^{-3/2}$ constant with
increasing $\varepsilon_{\alpha}$.

\begin{table}[t!]
\begin{tabular}{cccccccc}
$N$ & $\nu$ & $\alpha$ & 
$\ell_{f}/\ell_{d}$ & 
$\varepsilon_{\alpha}/\varepsilon_{I}$ & $\varepsilon_{\nu}/\varepsilon_{I}$ & 
$\eta_{\alpha}/\eta_{I}$ & $\eta_{\nu}/\eta_{I}$ \\ \hline
$2048$ & $2\times10^{-5}$ & $0.015$ & 
$26.2$ &
$0.54$ & $0.46$ &
$0.03$ & $0.97$ \\
$4096$ & $5\times10^{-6}$ & $0.024$ & 
$52.3$ &
$0.82$ & $0.18$ &
$0.08$ & $0.92$ \\
$8192$ & $2\times10^{-6}$ & $0.025$ & 
$80.5$ &
$0.92$ & $0.08$ &
$0.10$ & $0.90$ \\
$16384$ & $1\times10^{-6}$ & $0.03$ & 
$114.2$ &
$0.95$ & $0.05$ &
$0.12$ & $0.88$ \\ \hline
\end{tabular}
\caption{Parameters of the simulations. $N$ spatial resolution, 
$\nu$ viscosity, $\alpha$ friction, $L$ box size, $\ell_{f}=L/100$
forcing scale, $\ell_{d}=\nu^{1/2}/\eta_{\nu}^{1/6}$ enstrophy
dissipative scale, 
$\varepsilon_{I}$ energy injection rate, $\varepsilon_{\nu}$
viscous energy dissipation rate, $\varepsilon_{\alpha}$
friction energy dissipation rate,
$\eta_{I}$ enstrophy injection rate, $\eta_{\nu}$
viscous enstrophy dissipation rate, $\eta_{\alpha}$
friction enstrophy dissipation rate.}
\label{table1}
\end{table}

\begin{figure}[h!]
\includegraphics[draft=false,scale=0.68]{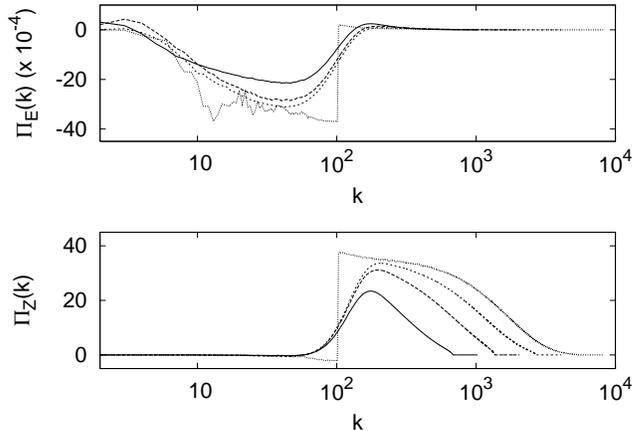}
\caption{Energy (top) and enstrophy (bottom) fluxes in Fourier space
at resolutions $2048$, $4096$, $8192$ and $16384$. Resolution
increases from smaller to larger absolute values of the minimum/maximum.
At resolution $16384$ statistics is computed on a single
frame.}
\label{fig1}
\end{figure}

In Figure~\ref{fig1} we plot the fluxes of energy and enstrophy 
in wavenumber space.
Observe that because change the resolution keeping the ratio $L/\ell_{f}$ 
constant, the only effect of reducing the resolution on the inverse 
cascade is the decrease of the energy transfered to large scales
(being $\varepsilon_{\alpha}=\varepsilon_{I}-\varepsilon_{\nu}$
with $\varepsilon_{\nu}$ proportional to $\nu$) while the extension
of the inertial range is almost unaffected.
The qualitative difference at $k \simeq k_{f}$ for the run
at $N=16384$ is due to the different forcing implemented in this case,
while statistical fluctuations are a consequence of the short time statistics.
These results confirm the robustness of the energy inertial range
regardless of the viscous dissipative scale, a further justification 
of many simulations of the inverse cascade
in which, because of the limited resolution, the forcing scale is
very close to the dissipative scale.

At variance with the inverse cascade, the direct enstrophy cascade
is strongly affected by finite resolution effects.
This is not a surprise because, by keeping $\ell_{f}$ fixed, 
the extension of the direct cascade is
simply proportional to $N$.
As it is shown in Figure~\ref{fig1}, we observe a range of wavenumbers
with almost constant flux $\Pi_{Z}(k)$ only for the runs at $N\ge 8192$.

\begin{figure}[h!]
\includegraphics[draft=false,scale=0.68]{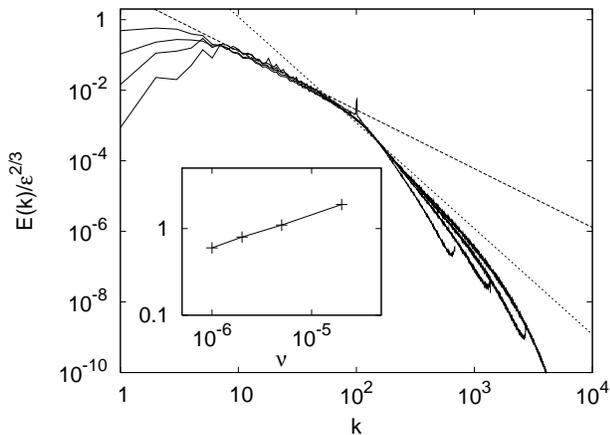}
\caption{Energy spectra for the two simulations for the different
resolutions (resolution increases with the extension of the tail).
Dashed and dotted lines represent the two predictions $C k^{-5/3}$ with $C=6$
and $k^{-3}$ respectively. Inset: correction to the Kraichnan exponent
$-3$ measured in the range $100 \le k \le 400$ as a function
of viscosity.}
\label{fig2}
\end{figure}

Figure~\ref{fig2} shows the energy spectra computed for the different
runs. We remark again that the only effect of finite resolution 
on the inverse cascade is the reduction of the energy transfered
to large scales, while the Kolmogorov scaling $k^{-5/3}$ is always 
observed with a Kolmogorov constant $C \simeq 6$ \cite{BCV00}
virtually independent on resolution.
The effect of finite resolution is, of course, more dramatic on the
enstrophy cascade range. We observe here a significative correction
to the Kraichnan spectrum $k^{-3}$  even for the $16384$ run, where
measure a scaling exponents close to $-3.6$.
We remark a similar steepening of the spectrum has been observed even 
for simulations with a more resolved direct cascade range
(here we have $k_{max}/k_f \simeq 55$ at the highest resolution). 
Despite these difficulties, there is a clear indication 
that the correction to the exponent is a finite-size effect which
eventually disappears by increasing the extension of the inertial
range (see inset of Fig.~\ref{fig2}).
The conclusion of these considerations is that a $k^{-3}$ spectrum
in stationary solutions of (\ref{eq:1}) could be achieved only by 
taking simultaneously the limits $L/\ell_{f} \to \infty$ and
$\ell_{f}/\ell_{d} \to \infty$ (i.e. for vanishing $\alpha$ and $\nu$).

A better understanding of the physical mechanism of the
cascades can be obtained by looking at the distribution
of fluxes in space. This can be obtained by using a filtering
procedure recently introduced and applied separately to the direct 
\cite{CEEWX03} and to the inverse \cite{CEERWX06} cascades. 
Thanks to the resolution of the present simulations, we are
able to analyze jointly both the cascade and the 
correlations among them.
Following \cite{CEEWX03}, we introduce a large-scale vorticity
field $\omega_{r}\equiv G_{r} \star \omega$ obtained from
the convolution of $\omega$ with a Gaussian filter $G_{r}$, 
and a large-scale velocity field
${\bm v}_{r}\equiv G_{r} \star {\bm v}$.
From these definitions, balance equations for the large-scale 
energy $e_{r}({\bm x},t)=1/2 |{\bm v}_r|^2$ and enstrophy 
$z_{r}({\bm x},t)=1/2 \omega_r^2$
densities are easily written (with a compact notation):
\begin{equation}
\partial_t (e_r,z_r) + {\bm \nabla} \cdot {\bm J}^{(e,r)}_r =
- \Pi^{(e,r)}_r - D^{(e,r)}_r + F^{(e,r)}_r 
\label{eq:2} 
\end{equation}
where ${\bm J}^{(e,z)}_r$ represent the transport of large-scale
energy and enstrophy, $D^{(e,z)}_r$ and $F^{(e,r)}_r$ represent
the large-scale dumping and forcing.
The energy/enstrophy fluxes $\Pi^{(e,z)}_r({\bm x},t)$,
representing the local transfer of energy/enstrophy from large scale
to scales smaller than $r$, are given by:
\begin{eqnarray}
\Pi^{(e)}_r({\bm x},t) &\equiv& - ({\tau}_{\alpha \beta})_{r} \nabla_{\alpha}
(v_{\beta})_{r} \label{eq:3} \\
\Pi^{(z)}_r({\bm x},t) &\equiv& - ({\sigma}_{\alpha })_{r} \nabla_{\alpha}
\omega_{r} \label{eq:4} 
\end{eqnarray}
where $({\tau}_{\alpha \beta})_{r}=(v_{\alpha} v_{\beta})_r - 
(v_{\alpha})_r (v_{\beta})_r$ and
$({\sigma}_{\alpha})_{r}=(v_{\alpha} \omega)_r - 
(v_{\alpha})_r \omega_r$.

\begin{figure}[h!]
\includegraphics[draft=false,scale=0.68]{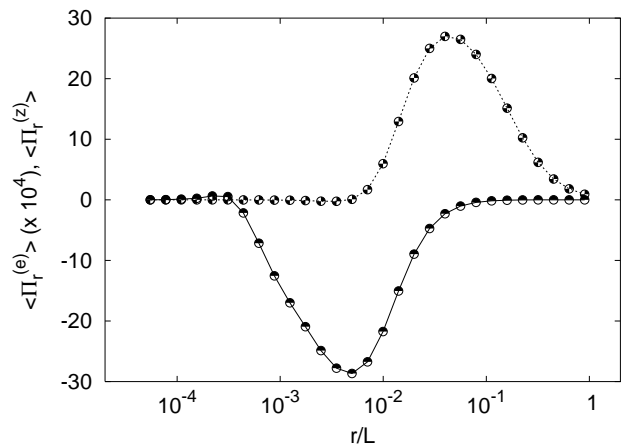}
\caption{Average energy (lower, continuous) and enstrophy (upper, dotted) 
fluxes in physical space at resolution $N=8192$. The energy flux
is multiplied by a factor $k_f^2=10^4$ for better visualization.}
\label{fig3}
\end{figure}

Fluxes (\ref{eq:3}-\ref{eq:4}) are expected to have a non-zero 
spatial mean in the inertial range of scales of irreversible turbulent
cascades: in particular a mean negative energy flux for $r>\ell_f$ 
(inverse cascade) and a mean positive enstrophy flux for 
$r < \ell_f$ (direct cascade).
Figure~\ref{fig3} shows the averaged physical fluxes 
$\langle \Pi^{(e,z)}_{r} \rangle$ as functions of the scale $r$.
The two cascades are evident, although the range of constant
flux is reduced with respect to the spectral case in both the
cascades (see Fig.~\ref{fig1}).

\begin{figure}[h!]
\includegraphics[draft=false,scale=0.23]{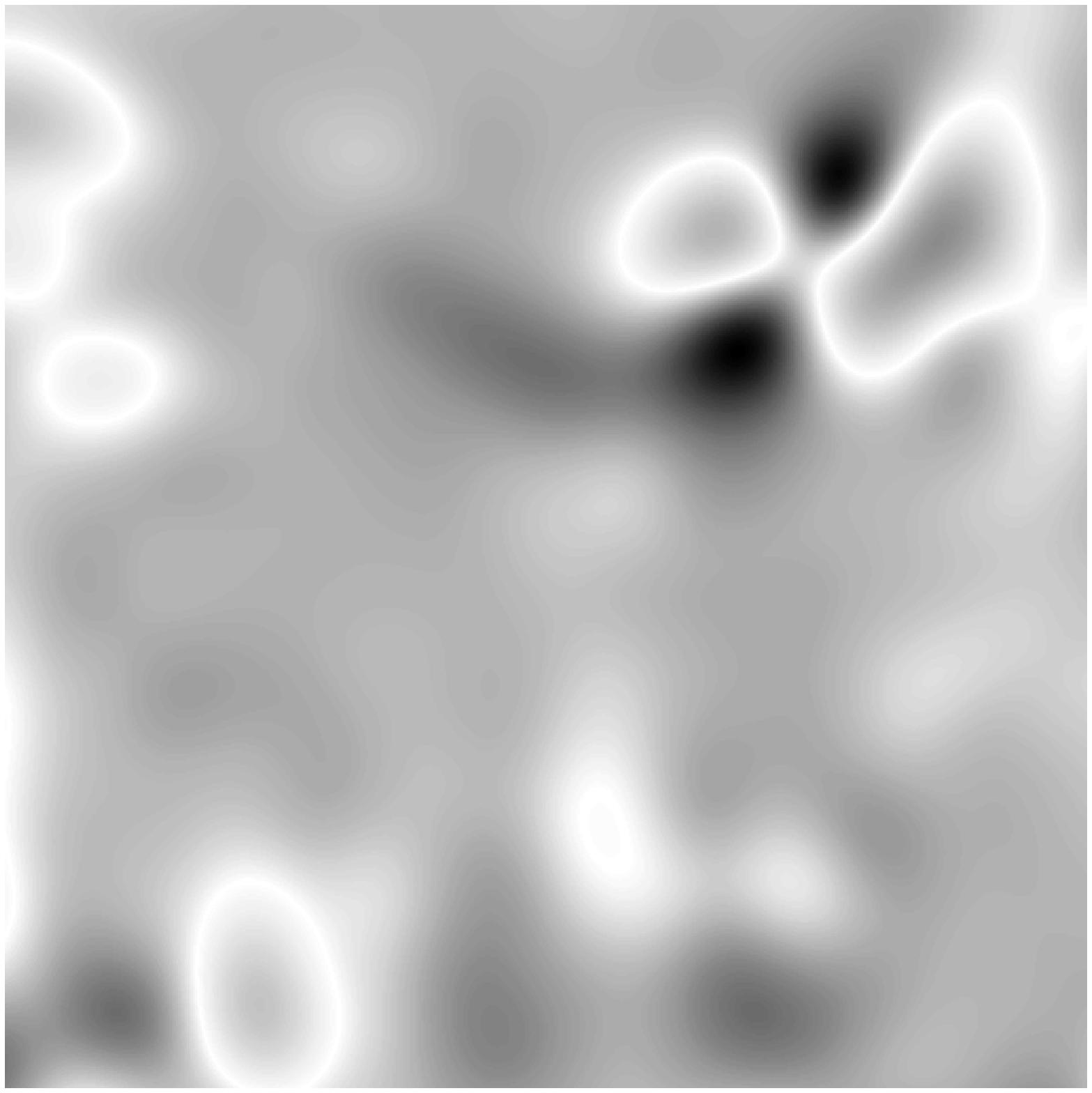}
\includegraphics[draft=false,scale=0.23]{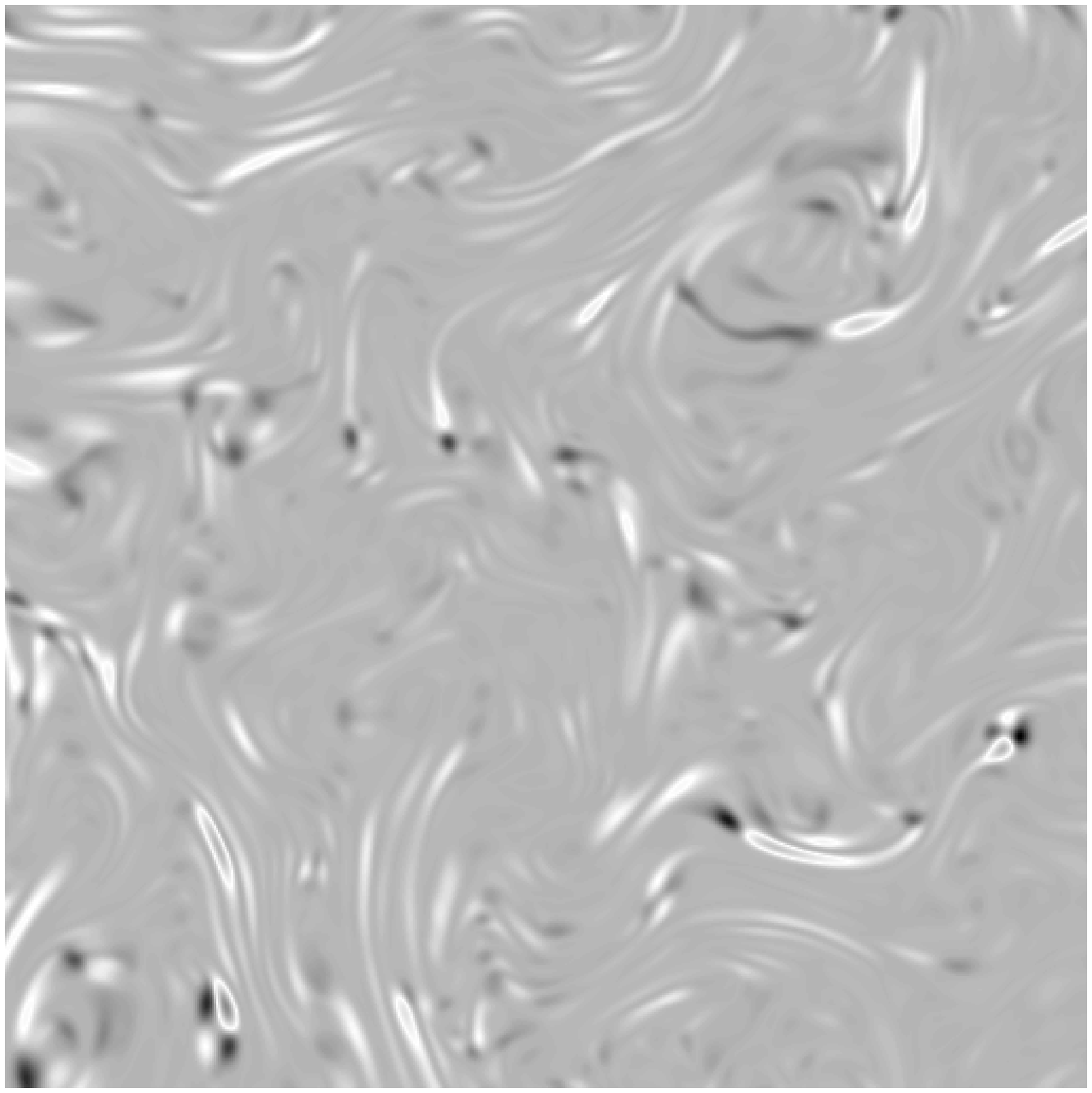}
\caption{Snapshot of energy (left) and enstrophy (right)
fluxes from the same vorticity field. Energy flux (\ref{eq:3}) is computed
at scale $r_1=0.025 L$ and enstrophy flux at $r_2=0.0025 L$, roughly 
corresponding to the minimum and maximum in Fig.~\ref{fig3}.
White and black correspond to positive and negative values respectively.}
\label{fig4}
\end{figure}

Local fluxes are strongly inhomogeneous in physical space:
there are relatively small regions of intense (positive and 
negative) flux in both the energy and enstrophy inertial ranges.
Figure~\ref{fig4} shows two snapshots of the energy and enstrophy
fluxes, computed from the same field at two different scales 
$r_1$ and $r_2$ corresponding to the minimum of energy flux
and the maximum of enstrophy flux respectively (see Fig.~\ref{fig3}).
An interesting information, obtained from Fig.~\ref{fig4} at a 
qualitative level, is that the most intense energy and enstrophy 
fluxes appear on different physical region without any apparent 
correlation.

\begin{figure}[h!]
\includegraphics[draft=false,scale=0.68]{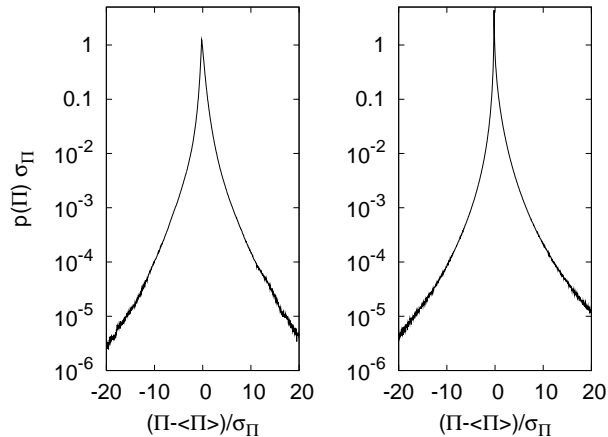}
\caption{Probability density function of energy 
$\Pi^{(e)}_{r_1}$ (left) and enstrophy $\Pi^{(e)}_{r_2}$ (right) 
fluxes normalized with their standard deviations. Resolution is
$N=8192$, filtering scales are $r_1=0.025 L$ and $r_2=0.0025 L$.}
\label{fig5}
\end{figure}

Figure~\ref{fig4} shows that both positive and negative fluxes are 
observed: locally both energy and enstrophy can go to 
smaller or larger scales. 
Figure~\ref{fig5} shows the probability density functions of 
the two fluxes. As observed in previous studies \cite{CEEWX03,CEERWX06}, 
the shapes of pdf's are nearly symmetric, confirming the qualitative
picture inferred from Fig.~\ref{fig4}. The mean values in both cases
is the result of strong cancellations as its ratio with the standard 
deviation is $-0.22$ for the energy flux and $0.16$ for the enstrophy flux. 
The skewness is also small as it is about $0.9$ and $1.2$ for the energy 
and enstrophy fluxes respectively.

\begin{figure}[hbt]
\includegraphics[bb=65 80 390 250,draft=false,scale=0.73]{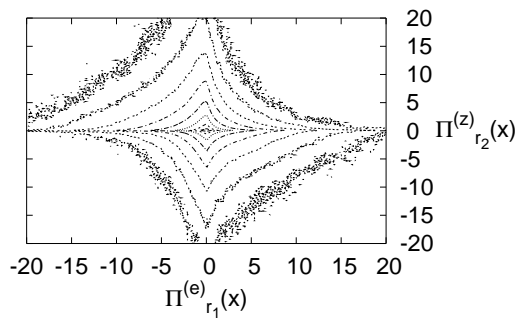}
\caption{Joint probability density function 
$p(\Pi^{(e)}_{r_1},\Pi^{(z)}_{r_2})$ of energy (x-axis) and
enstrophy (y-axis) fluxes. Scales $r_1$ and $r_2$ as in Fig.~\ref{fig5}.
Contours are plotted on a logarithmic
scale while fluxes are normalized with their standard deviations.}
\label{fig6}
\end{figure}

Figure~\ref{fig6} shows the joint probability density function
$p(\Pi^{(e)}_{r_1},\Pi^{(z)}_{r_2})$ for the same scales
$r_1$ and $r_2$.
This pdf is not far from the product of the marginal
distributions shown in Fig.~\ref{fig5}, a condition for independence.
Indeed, the cross correlation among $\Pi^{(e)}_{r_1}$ and
$\Pi^{(z)}_{r_2}$ is only $C(r_1,r_2) \simeq -0.15$.
Of course, it is very different if we consider the correlation
at the same scale, for which we find $C(r,r) \simeq 1$ despite
the fact that one of the fluxes has zero mean.
A possible interpretation of the observed small value of correlation
is in favor of the classical picture of independence of the 
two cascades which is here obtained at a local level.
Therefore, our result is, a posteriori, a support to many
experimental and numerical studies in which, due to finite size
constraints, only one cascade is realized.

In conclusion, we have presented statistical analysis of 
high resolution direct numerical simulations of 2D Navier-Stokes
equations which clearly reproduce, for the first time,
the double cascade scenario predicted by Kraichnan almost $40$ years ago.

\begin{acknowledgments}
Simulations were performed on the IBM-CLX cluster of Cineca (Bologna,
Italy) and on the {\it Turbofarm} cluster at the INFN computing
center in Torino.
\end{acknowledgments}                              
                             


\end{document}